\documentclass[10pt,aps,prl,twocolumn,superscriptaddress]{revtex4-2} 
\usepackage[utf8]{inputenc}
\usepackage[T1]{fontenc}
\usepackage{amssymb}
\usepackage{color}
\usepackage{graphicx}
\usepackage[hidelinks]{hyperref}
\usepackage{nameref}
\usepackage{amsmath}
\begin{document} 
\title{\texorpdfstring{Removal of $K$-mixing in angular momentum projected nuclear wave functions}{Removal of K-mixing in angular momentum projected nuclear wave functions}}

\author{Xiao Lu}
\email[]{luxiao@itp.ac.cn} 
\affiliation{CAS Key Laboratory of Theoretical Physics, Institute of Theoretical Physics, Chinese Academy of Sciences, Beijing 100190, China}
\author{ Zhan-Jiang Lian}
\email[]{lianzhanjiang@cnncmail.cn} 
\affiliation{China Institute of Atomic Energy, Beijing 102413, China}
\author{ Zao-Chun Gao}
\email[]{zcgao@ciae.ac.cn} 
\affiliation{China Institute of Atomic Energy, Beijing 102413, China}

\date{\today}

\begin{abstract}
Angular momentum projection plays a key role in studying quantum many-body systems with rotational invariance such as atomic nuclei. At a given spin $J$, one can generate $2J+1$ angular momentum projected states labeled with $-J\leq K \leq J$ from a deformed Slater determinant. Usually, a nuclear wave function with $K$-mixing can be expressed as a superposition of all these $2J+1$ projected states,  where the coefficients can be obtained by solving the generalized eigenvalue equation. In this Letter, we report a new fundamental feature that the frequently discussed $K$-mixing in the angular momentum projected nuclear wave function can be safely removed. Strikingly, we found that such nuclear wave function with $K$-mixing can always be equivalently replaced by a single projected state with any given $K$. Consequently, such nuclear wave function can be significantly simplified, especially for high-spin states. This also indicates that the $K$-mixing in the angular momentum projected nuclear wave functions, adopted by many present-day nuclear models, does not carry any physical meaning, and is essentially different from that $K$-mixing caused by the Coriolis force in the cranked shell model.
\end{abstract}

%\keywords{}
\maketitle 

Mean-field approaches has been widely applied in the fields of physics and chemistry, providing a good qualitative understanding of the quantum many-body systems \cite{PR80,chaikin1995principles,parr1995density}. However, one of the most salient characteristics of mean-field approaches is that the solutions often spontaneously break symmetries of the Hamiltonian, making it impossible to label the system with symmetry quantum numbers such as angular momentum, parity, etc. To provide a more detailed and accurate characterization of many-body systems, such as the spectroscopy of the atoms, molecules or nuclei, it is necessary to restore symmetries, particularly rotational symmetry.

Angular momentum projection is considered to be a very powerful technique in quantum many-body problem calculations. By effectively removing the spurious, symmetry-breaking components from the mean-field wave function, this approach isolates the state with a well-defined angular momentum quantum number $J$. The remaining state, commonly referred to as the angular momentum projected or simply the projected state, is particularly suitable for the construction of wave functions defined in the laboratory frame of reference. Reflecting its indispensable utility, angular momentum projection is now extensively integrated into various advanced methodologies employed in quantum many-body physics and chemistry, including the projected shell model \cite{SY95,Sunyang16}, variation after projection (VAP) methods \cite{VAMPIR04,Egido10,tuya17,vap18,VAP22,Shimizu21,2021SA,PRB04,PRB05,PRB13}, stochastic quantum Monte Carlo approaches \cite{otsuka2001monte,Shimizu12,Shimizu17}, and many other sophisticated beyond-mean-field approaches \cite{Bender03,2006prc,Yao09,Yao11,Bally14,Bally21,Bally2024,SRMF21,Che2023,wang22,rong23,sun2021,sun21,zhao16,wangyk22,wangyk24,WANG2024,yyx22,satula2014simple,bkaczyk2021mirror}. 
 
Generally, with a given spin $J$, one can construct $2J+1$ projected states from a mean-filed wave function $|\Phi\rangle$. These projected states are labeled by $K$, which originates from the quantum number of the operator $\hat{J}_3$, the third component of the angular momentum but defined in the intrinsic frame of reference.  Usually, all these $2J+1$ projected states, $\hat{P}^J_{MK}|\Phi\rangle$, are used to form a many-body trial wave function written as
\begin{eqnarray}\label{psi}
|\Psi_{JM}\rangle=\sum_K f_K \hat{P}^J_{MK}|\Phi\rangle.
\end{eqnarray}
This naturally introduces the concept of $K$-mixing, meaning that the many-body wave function contains the $\hat{P}^J_{MK}|\Phi\rangle$ states with different $K$. Indeed, in almost all modern nuclear theories involving angular momentum projection, $K$-mixing is believed to be necessary to obtain a lower energy of the nuclear system. However, using such a wave function can lead to computational challenges or problems in the calculations. On the one hand, the angular momentum projected states are, in general, non-orthogonal to each other. This often leads to an overcomplete model space, which in turn can result in singular energy kernels that cannot be properly integrated to restore symmetries \cite{satula2014simple,bkaczyk2021mirror}. Moreover, if the method involves VAP, such conditions become even more crucial and can even cause the collapse of the iteration process \cite{VAP22}. Clearly, the wave function composed of $K$-mixing projected states would further exacerbate these problems. Additionally, using the $K$-mixing wave function to calculate the high spin states of the many-body system will result in a very large computational cost, which is not conducive to efficient calculations.

It is worth noting that $K$-mixing in projected wave functions is not strictly necessary. As noted in Ref. \cite{otsuka2001monte}, the states of the same $J$ and $M$ but with different $K$'s are usually mixed in the eigenstates of the Hamiltonian. Typically, only one or a few linear combinations of such states of different $K$ values are important and are kept as the Quantum Monte Carlo Diagonalization basis vectors. Particularly, in Ref. \cite{VAP22}, we have made an interesting comparison of two different wave functions which are independently optimized in the framework of the VAP method. The first wave function is $|\Psi_{JM}\rangle$ given in Eq. (\ref{psi}) and the second one is simply a projected state $\frac{\hat{P}^J_{MK}|\Phi\rangle}{\sqrt{\langle\Phi|P^J_{KK}|\Phi\rangle}}$ with $K$ randomly chosen. These two wave functions look quite different, but when they are fully optimized, their corresponding energies are very close to each other, and their overlap is very close to 1. In other words, they are essentially equivalent to each other.  If we consider the many-body system as a rotating nucleus, this equivalence can be understood through the following explanation. At a given spin $J$, when the optimized mean-field state (Hartree-Fock or Hartree-Fock-Bogoliubov state) $|\Phi\rangle$ and the corresponding energy $E^{J}_1$ are obtained with the first wave function $|\Psi_{JM}\rangle$ in the VAP calculation, the orientation of rotational axis relative to $|\Phi\rangle$ for $E^{J}_1$ is determined. Since $E^{J}_1$ is the lowest energy, the moment of inertia along this rotational axis should be the largest. Alternatively, one can first fix the orientation of the rotational axis relative to $|\Phi\rangle$ and then perform energy minimization, as is done in the VAP calculation with the second wave function. It is expected that the moment of inertia along this fixed rotational axis will reach the same maximum as that for $E^{J}_1$, leading to a convergence of the energy $E^J(K)$ calculated with the second wave function to $E^{J}_1$.
% This fact tells us that $K$-mixing is unnecessary in the variation after projection (VAP) methods. 
Indeed, in several later VAP calculations \cite{Lian22,Lian23,Xiao23prc,Xiao23cpc,lian2024}, the wave functions without $K$-mixing are taken, and good shell model approximation can be achieved. 

%\red{Actually, effectively describing the atomic nucleus is a central issue, where one of the primary challenges lies in the high computational burden. Moreover, constructing appropriate nuclear wave functions has always been an unresolved issue. Obviously}
Clearly, without $K$-mixing, the many-body wave function can be written in a more compact form and the associated many-body calculations become much easier. This is especially helpful in the studies of the high-spin states \cite{VAP22}. Thus it is of important significance to investigate if an arbitrary $|\Psi_{JM}\rangle$ can be simplified as a single projected state.

In this Letter, we first analytically prove that any $|\Psi_{JM}\rangle$ wave function can indeed be transformed into a single projected state generated from a reference state $|\phi\rangle$ that does not involve $K$-mixing. This confirms the $K$-mixing in any angular momentum projected wave functions, whether optimized or not, can be safely removed. Such $|\phi\rangle$ state looks very complicated. But interestingly, practical calculations show that $|\phi\rangle$ can be as simple as a single Slater Determinant (SD).

Let us first transform the $K$-mixing wave function $|\Psi_{JM}\rangle$ into a projected state without $K$-mixing. According to the theory of angular momentum, it is known that
\begin{eqnarray}
\hat{J}_{\pm}|\nu JK\rangle = \hbar\sqrt{(J\mp K)(J\pm K+1)}|\nu JK\pm 1\rangle
\end{eqnarray}
and its Hermitian conjugate
\begin{eqnarray}\label{lad}
\langle\nu JK|\hat{J}_{\mp}= \hbar\sqrt{(J\mp K)(J\pm K+1)}\langle\nu JK\pm 1|,
\end{eqnarray}
where $\nu$ refers to other quantum numbers. By definition, we also have \cite{PR80,varshalovich1988quantum}
\begin{eqnarray}\label{pjmk}
\hat{P}^J_{MK}=\sum_\nu |\nu JM\rangle\langle\nu JK|.
\end{eqnarray}
Combining Eq. (\ref{lad}) and Eq. (\ref{pjmk}), one can immediately have
\begin{eqnarray}\label{pjlad1}
\hat{P}^J_{MK+1}&=& c_{K+1} \hat{P}_{M K}^J \hat{J}_{-},\\
\hat{P}^J_{MK-1}&=& c_{K} \hat{P}_{M K}^J \hat{J}_{+},
\end{eqnarray}
where
\begin{eqnarray}\label{ck}
c_{K}\equiv\frac{1}{\hbar\sqrt{(J+K)(J-K+1)}}.
\end{eqnarray}
Thus for an arbitrary quantum state $|\Phi\rangle$, we have
\begin{eqnarray}\label{pju}
\hat{P}^J_{MK+1}|\Phi\rangle&=& \hat{P}_{M K}^J \left(c_{K+1} \hat{J}_{-}|\Phi\rangle\right),\\
\hat{P}^J_{MK-1}|\Phi\rangle&=& \hat{P}_{M K}^J \left(c_{K} \hat{J}_{+}|\Phi\rangle\right).\label{pjd}
\end{eqnarray}
Note that the ladder operators $\hat{J}_{\pm}$ are defined in the intrinsic frame of reference.
From Eqs. (\ref{pju}) and (\ref{pjd}), $K$ is not a good quantum number of the wave function in the laboratory frame of reference. By recursively using Eqs. (\ref{pju}) and (\ref{pjd}), one can get
\begin{eqnarray}\label{pjlad}
\hat{P}^J_{MK+2}|\Phi\rangle&=&\hat{P}_{M K}^J\left( c_{K+2}c_{K+1}(\hat{J}_{-})^2|\Phi\rangle\right),\\
\hat{P}^J_{MK-2}|\Phi\rangle&=&\hat{P}_{M K}^J\left( c_{K-1}c_{K}(\hat{J}_{+})^2|\Phi\rangle\right),
\end{eqnarray}
and so on. Therefore, one can write down
\begin{equation}\label{pjkpro}
|\Psi_{JM}\rangle=\sum_{K'=-J}^{J}f_{K'}\hat{P}_{M K'}^J |\Phi\rangle=\hat{P}_{M K}^J |\phi\rangle,
\end{equation}
where
 \begin{eqnarray}\label{phi1}
    |\phi\rangle &=&\sum_{K'=K+1}^J\left(f_{K'}\left(\prod_{i=K+1}^{K'}c_i\right)(\hat{J}_{-})^{K'-K}+f_K\right.\nonumber\\
    &&\left.+\sum_{K'=-J}^{K-1}f_{K'}\left(\prod_{i=K'+1}^{K}c_i\right)(\hat{J}_{+})^{K-K'}\right)|\Phi\rangle.
    \end{eqnarray}
 
Equation (\ref{pjkpro}) clearly tells us that any angular momentum projected wave functions with $K$-mixing in Eq. (\ref{psi}) can, in principle, be replaced by a projected state generated from a reference state $|\phi\rangle$, which is characterized by a given $K$.
But, the state $|\phi\rangle$ in Eq. (\ref{phi1}) looks so complicated that one cannot see any advantage of using such $|\phi\rangle$ in practical calculations. However, we have learned that the projection is invertible, and two quite difference reference states may correspond to one identical projected state \cite{VAP15}. This implies that there may exist a simple state (for instance, a SD) $|\Phi'\rangle$ to take the place of $|\phi\rangle$ in Eq. (\ref{pjkpro}).
An immediate case of Eq. (\ref{pjkpro}) is the following identity \cite{PR80}
\begin{eqnarray}\label{pjr}
\hat{P}^J_{MK}\hat R(\Omega)|\Phi\rangle=\sum_{K'=-J}^J D^J_{KK'}(\Omega)\hat{P}^J_{MK'}|\Phi\rangle.
\end{eqnarray} 
Here, $\hat R (\Omega)$ is the rotation operator and $D^{J}_{MK}(\Omega)=\langle JM|\hat R (\Omega)|JK\rangle$ is the Wigner $D$-function. With $|\Phi\rangle$ being a SD, the state $|\Phi'\rangle=\hat R(\Omega)|\Phi\rangle$ is a SD as well.

Interestingly, if $|\Phi\rangle$  is axial, and we assume $\hat J_3|\Phi\rangle=K_0|\Phi\rangle$, then Eq. (\ref{pjr}) can be reduced to
\begin{eqnarray}\label{pjra}
\hat{P}^J_{MK}\hat R(\Omega)|\Phi\rangle= D^J_{KK_0}(\Omega)\hat{P}^J_{MK_0}|\Phi\rangle.
\end{eqnarray}
This means that all projected states $\hat{P}^J_{MK}\hat R(\Omega)|\Phi\rangle$ with different $K$ are exactly identical, and are the same as $\hat{P}^J_{MK_0}|\Phi\rangle$. Thus, it suffices to pick just one of these states, either $\hat{P}^J_{MK}\hat R(\Omega)|\Phi\rangle$ or $\hat{P}^J_{MK_0}|\Phi\rangle$, to construct the same nuclear wave function. The corresponding examples have been explicitly demonstrated in Refs. \cite{Yao09,Egido10}. This again confirms that the $K$ number in $\hat{P}^J_{MK}|\Phi\rangle$ has no physical meaning.

Unfortunately, it is not clear if there always exists a projected SD, $\hat{P}^J_{MK}|\Phi'\rangle$, which is exactly identical to a $K$-mixing $|\Psi_{JM}\rangle$ wave function. However, in numeric calculations, one can investigate if $\hat{P}^J_{MK}|\Phi'\rangle$ can sufficiently approach to the $|\Psi_{JM}\rangle$ state.
Indeed, such calculations have been implemented in our previous VAP calculations \cite{VAP22}. As introduced above, the fully optimized VAP wave function in Eq. (\ref{psi}) can always be replaced by a single projected state with any given $K$, generated from an SD without $K$-mixing. This inspires us to investigate if it is a general phenomenon that, for a given arbitrary $|\Psi_{JM}\rangle$, one can numerically find a SD $|\Phi'\rangle$ to approximately satisfy

\begin{equation}\label{pjk}
|\Psi_{JM}\rangle=\sum_{K'=-J}^{J}f_{K'}\hat{P}_{M K'}^J |\Phi\rangle\approx\frac{\hat{P}^J_{MK}|\Phi'\rangle}{\sqrt{\langle\Phi'|P^J_{KK}|\Phi'\rangle}}
\end{equation}
with a high numerical precision. In the subsequent calculations, where a many-body system is considered as an atomic nucleus, one can find that such alternative SD can be naturally identified during the variational process. 
In order to make the above argument as general as possible, we choose various $|\Phi\rangle$ SDs in attempt to cover a wide range of nuclear states.
To study the $K$-mixing, $|\Phi\rangle$ is assumed to be a non-axial and reflection asymmetric SD, in which the single particle states are obtained by diagonalizing the following deformed single particle Hamiltonian
\begin{eqnarray}\label{hsp}
\hat{H}_{s.p.}&=&h_0+\varepsilon_{3} \hbar\omega \rho^2\sqrt{\frac{4\pi}{7}}Y_{30}\nonumber\\
&&-\frac{2}{3}\hbar\omega\rho^2\varepsilon_2\sqrt{\frac{4\pi}{5}}[\cos\gamma Y_{20}\nonumber\\
&&-\frac{\sin\gamma}{\sqrt{2}}(Y_{22}+Y_{2-2})],
\end{eqnarray}
where, $\varepsilon_2$, $\gamma$ and $\varepsilon_{3}$ are the quadrupole, triaxial and octupole deformation parameters, respectively. $h_0$ is simply the one-body term of the adopted shell model Hamiltonian. For the oscillator frequency $\omega$, we simply take $\hbar\omega=41.0A^{-1/3}$ MeV and $A$ is the mass number. $\rho=\sqrt{\frac{m\omega}{\hbar}}r$ is dimensionless. Since $|\Phi\rangle$ is  reflection asymmetric, we put the parity projection into $|\Psi_{JM}\rangle$ and the generalized wave function $|\Psi_{J\pi M\alpha}\rangle$ can be written as
\begin{eqnarray}\label{wf}
|\Psi_{J\pi M\alpha}\rangle=\sum_{K'=-J}^Jf^{\alpha}_{K'}\hat{P}^{J\pi}_{MK'}|\Phi\rangle.
\end{eqnarray}
Here, $\hat{P}^{J\pi}_{MK}=\hat{P}^J_{MK}\hat{P}^\pi$ and $\hat{P}^\pi=\frac12(1+\pi\hat P)$ is the parity projection operator with $\pi=\pm1$. $\alpha$ is used to label the states with the same $J$, $\pi$ and $M$. If $f^{\alpha}_{K'}$ in Eq. (\ref{wf}) happens to be $D^J_{KK'}(\Omega)$, then $|\Psi_{J\pi M\alpha}\rangle$ can be simply compacted into a single projected state according to Eq. (\ref{pjr}). However, in practical calculations, the $f^{\alpha}_{K'}$ coefficients and the corresponding energy $E^{J\pi}_\alpha$ are determined by solving the following generalized eigenvalue equation of order $2J+1$,
\begin{eqnarray}\label{hw}
\sum_{K'=-J}^{J}(H^{J\pi}_{KK'}-E^{J\pi}_\alpha N^{J\pi}_{KK'})f^{\alpha}_{K'}=0,
\end{eqnarray}
where $H^{J\pi}_{KK'}=\langle\Phi|\hat H P^{J\pi}_{KK'}|\Phi\rangle$ and $N^{J\pi}_{KK'}=\langle\Phi|P^{J\pi}_{KK'}|\Phi\rangle$. For convenience, we assume $E^{J\pi}_1\leq E^{J\pi}_2\leq\cdots\leq E^{J\pi}_{2J+1}$. The $f^{\alpha}_{K'}$ coefficients are imposed to satisfy the normalization condition, so that $\langle\Psi_{J\pi M\alpha}|\Psi_{J\pi M\alpha}\rangle=1$.

Here, we are only interested in the lowest state ($\alpha=1$). To find a single projected SD $\hat{P}^{J\pi}_{MK}|\Phi'\rangle$ as a substitute for $|\Psi_{J\pi M1}\rangle$, one needs to calculate the overlap between $|\Psi_{J\pi M1}\rangle$ and $\hat{P}^{J\pi}_{MK}|\Phi'\rangle$ which is defined as
\begin{equation}\label{OK}
O_K=\frac{|\langle\Psi_{J\pi M1}|\hat{P}^{J\pi}_{MK}|\Phi'\rangle|}{\sqrt{\langle\Phi'|P^{J\pi}_{KK}|\Phi'\rangle}},
\end{equation}
where $|\Phi'\rangle$ is the SD that needs to be varied so that $O_K$ can be maximized. Ideally, if $O_K=1$, then  $\hat{P}^{J\pi}_{MK}|\Phi'\rangle$ is exactly equivalent to $|\Psi_{J\pi M1}\rangle$. In the present work, we try to demonstrate a very interesting phenomenon: the maximum of $O_{K}$ are always very close to 1 in our randomly selected examples. 
%In this case, the corresponding $|\Phi'\rangle$ is referred to as the variational SD.

In the practical calculations, we equivalently minimize $R_K$, the reciprocal of $O^2_K$,
\begin{equation}\label{rk}
R_K=\frac{1}{O^2_K}=\frac{\langle\Phi'|P^{J\pi}_{KK}|\Phi'\rangle}
{\langle\Phi'|\hat{P}^{J\pi}_{KM}|\Psi_{J\pi M1}\rangle\langle\Psi_{J\pi M1}|\hat{P}^{J\pi}_{MK}|\Phi'\rangle},
\end{equation}
 rather than maximizing $O_K$, so that the optimization algorithm in our VAP calculation can be directly applied. The gradient and the Hessian matrix of $R_K$ with respect to the variational parameters can be easily obtained using the available matrix elements required in our VAP calculation. This is actually a new VAP calculation but for the minimization of $R_K$.

\begin{figure}
 \centering
 \includegraphics[width=8cm]{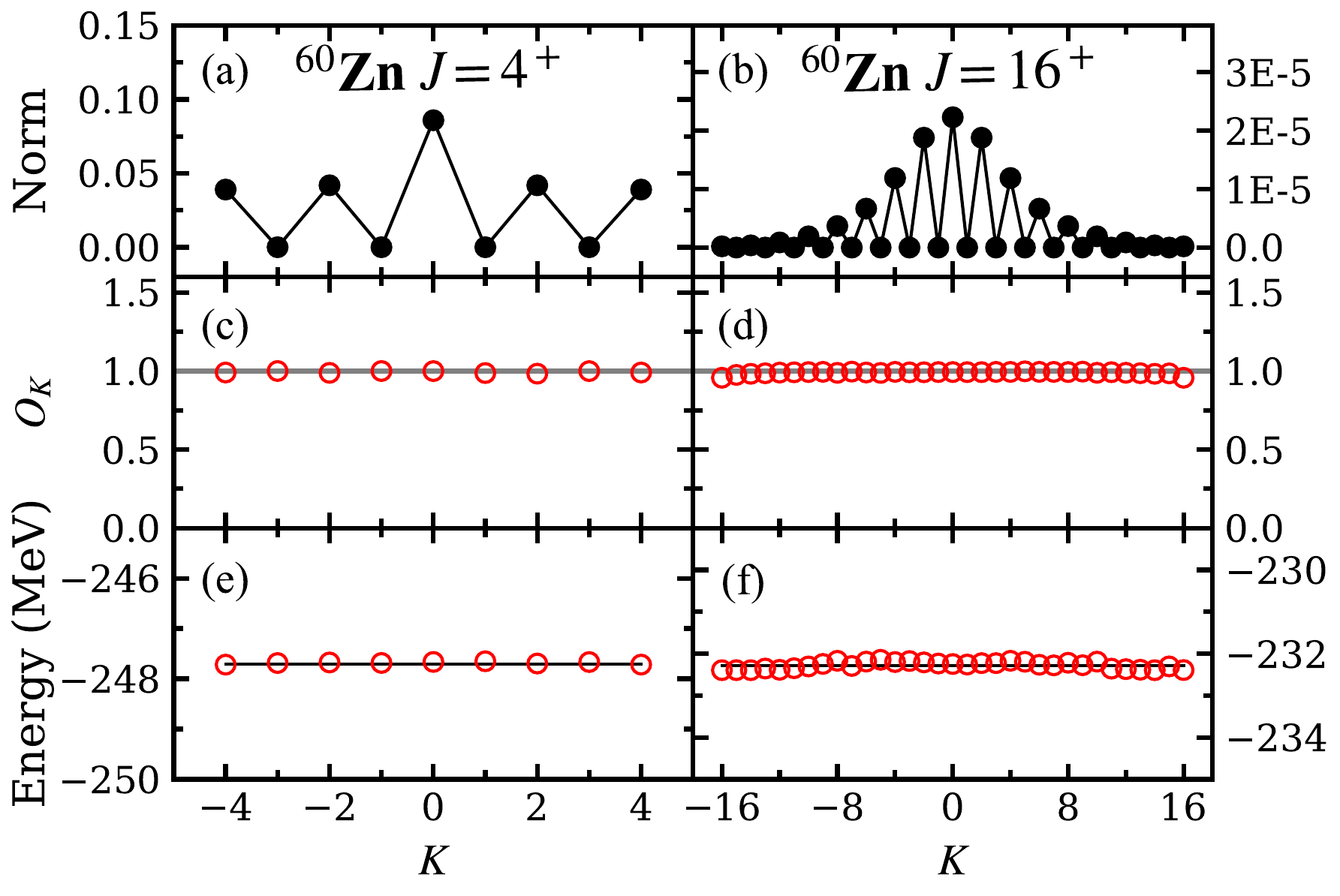}
 \caption{\label{Zn60}(Color online) Upper panel: The $\langle \Phi|P^{J\pi}_{KK}|\Phi\rangle $ norms of the projected states with (a) $J^\pi=4^+$ and (b) $J^\pi=16^+$, which are generated from the deformed SD in $^{60}$Zn and are used to form the $|\Psi_{J\pi M1}\rangle$ wave function. The single particle states in the SD are calculated from Eq. (\ref{hsp}) with $\varepsilon_2=0.3$ and $\gamma=40^\circ$. Middle panel: The corresponding $O_K$ overlaps for (c) $J=4^+$ and (d) $J=16^+$. A grey line is drawn to guide the eye. Lower panel: The energies of the calculated $|\Psi_{J\pi M1}\rangle$ (black line) and the equivalent $P^{J\pi}_{MK}|\Phi'\rangle$ projected states (red circle) for (e) $J=4^+$ and (f) $J=16^+$, respectively. The GXPF1A interaction is adopted.
}
\end{figure}

\begin{figure} 
 \centering
 \includegraphics[width=8cm]{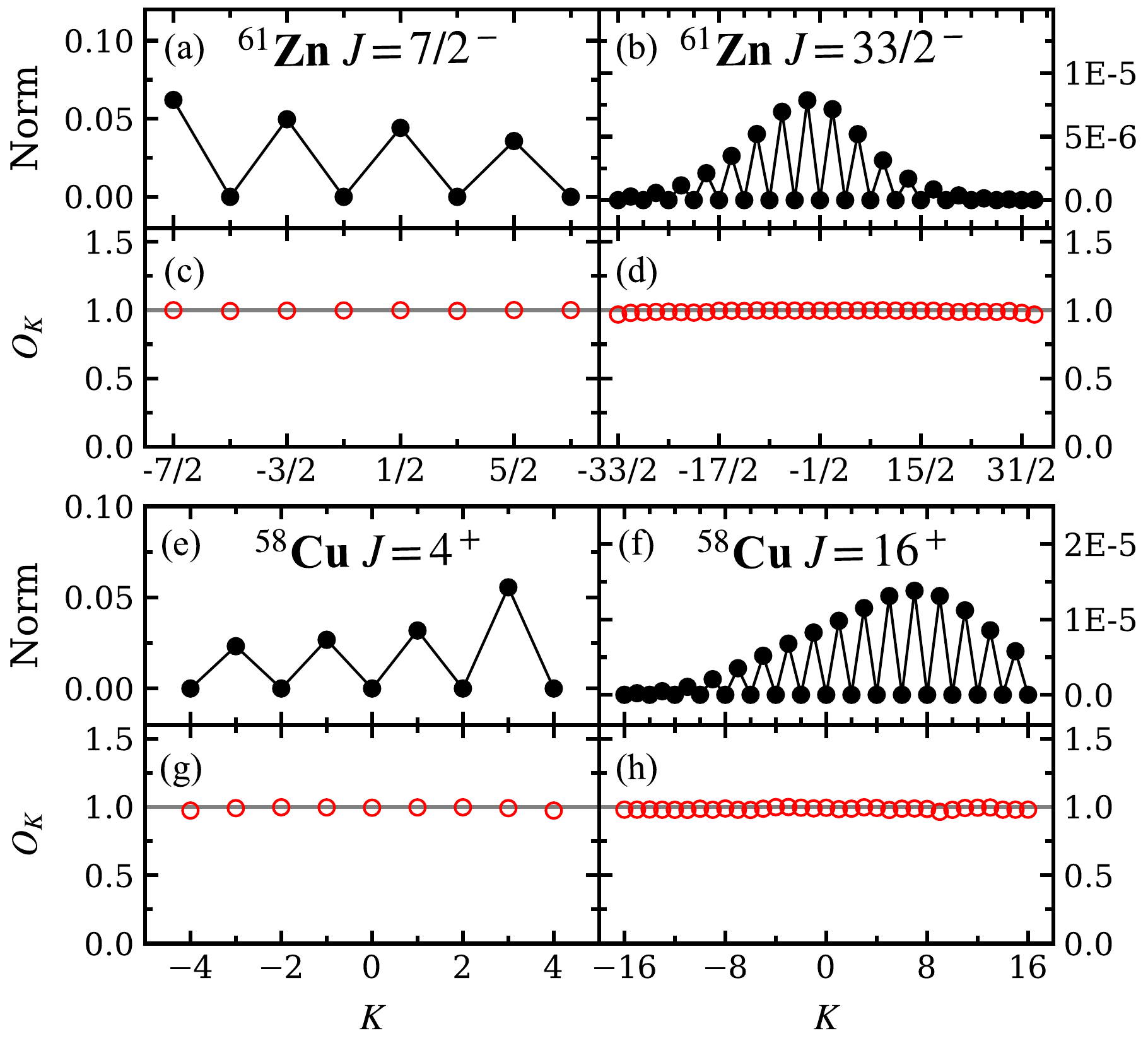}
 \caption{\label{zncu}(Color online) Same as Fig. \ref{Zn60}, but for odd-$A$ nucleus $^{61}$Zn and odd-odd nucleus $^{58}$Cu.
}
\end{figure}

Let us calculate the yrast $J^\pi=4^+$ and $16^+$ states in $^{60}$Zn within the $fp$ shell model space. For simplicity, we use the same $|\Phi\rangle$ for both $J^\pi=4^+$ and $16^+$ states. The single particle states in this $|\Phi\rangle$ are calculated from Eq. (\ref{hsp}) with $\varepsilon_2=0.3$ and $\gamma=40^\circ$, so that we know $|\Phi\rangle$ has a clear triaxial shape. The calculated $\langle \Phi|P^{J\pi}_{KK}|\Phi\rangle $ norms as functions of $K$ are shown in Fig. \ref{Zn60} (a) and (b). Then, $|\Psi_{J\pi M1}\rangle$ is obtained from Eq. (\ref{hw}) with $\hat H$ taken to be the GXPF1A interaction \cite{gxpf1a}. Put $|\Psi_{J\pi M1}\rangle$ into Eq. (\ref{rk}), one can minimize the $R_K$ value for each $K$ by varying the $|\Phi'\rangle$ SD. Then the corresponding maximum of $O_K$ can be obtained. As shown in Fig. \ref{Zn60} (c) and (d), all the calculated $O_K$ maxima for $K=-J,-J+1,\cdots,J$ are very close to 1, indicating the $|\Psi_{J\pi M1}\rangle$ with explicit $K$-mixing can be equivalently replaced by a single $\hat{P}^{J \pi} _{MK}|\Phi'\rangle$ projected state with $K$ arbitrarily chosen for both low spin states and high spin states. As an extra confirmation, one can calculate the energies of $\hat{P}^{J\pi}_{MK}|\Phi'\rangle$ states, and compare with the one for the $|\Psi_{J\pi M1}\rangle$ state. As shown in Fig. \ref{Zn60} (e) and (f), all the energy differences between $\hat{P}^{J \pi} _{MK}|\Phi'\rangle$ and $|\Psi_{J\pi M1}\rangle$ are within 100 keV. We also calculate the states in odd-$A$ $^{61}$Zn  and odd-odd $^{58}$Cu. The results are shown in Fig. \ref{zncu}. Again, we found the equivalent $\hat{P}^{J \pi} _{MK}|\Phi'\rangle$ projected SDs for those states in odd-$A$ and odd-odd nuclei.

\begin{figure} 
 \centering
 \includegraphics[width=8cm]{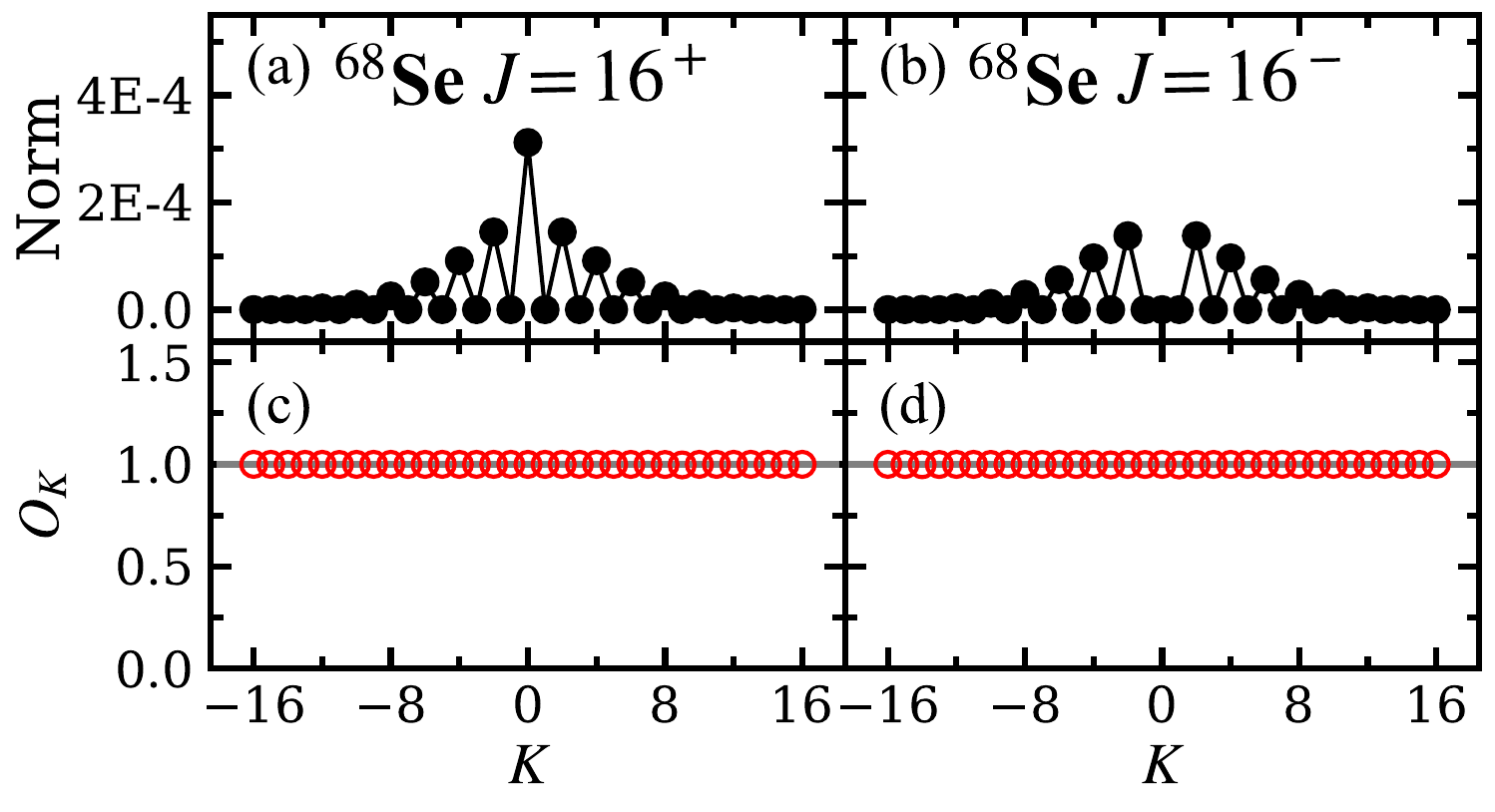}
 \caption{\label{Se68}(Color online) Same as Fig. \ref{Zn60}, but for $J=16$ in $^{68}$Se with both parities. The single particle states in the adopted $|\Phi\rangle$ SD are calculated from Eq. (\ref{hsp}) with $\varepsilon_2=0.45$,  $\gamma=40^\circ$ and $\varepsilon_{3}=0.3$. The JUN45 interaction is adopted.
}
\end{figure}

The above examples are in the $fp$ model space, so the parity projection is essentially unnecessary. In more general cases, $|\Phi\rangle$ can be both non-axial and reflection asymmetric. The parity projection should be performed.  In this instance, the wave functions of both $16^+$ and $16^-$ states in $^{68}$Se are studied in the $jj44$ model space, which constitutes of $f_{5/2}$, $p_{3/2}$, $p_{1/2}$ and $g_{9/2}$ orbits. The Hamiltonian is taken to be JUN45 \cite{jun45}. To generate a triaxial and octupole deformed SD, we simply take $\varepsilon_2=0.45$, $\gamma=40^\circ$ and $\varepsilon_{3}=0.3$. Then the ground state SD is projected onto spin $J=16$ with both parities. The $\langle \Phi|P^{J\pi}_{KK}|\Phi\rangle $ norms are shown in Fig. \ref{Se68} (a) and (b). Without exception, all the calculated $O_K$ values are still very close to 1 after the variation of $|\Phi'\rangle$. This example shows even for a more complicated $K$-mixing $|\Psi_{J\pi M1}\rangle$ with a reflection asymmetric SD, one can still find an equivalent single projected state with an SD to take the place of it.

\begin{figure}
 \centering
 \includegraphics[width=8cm]{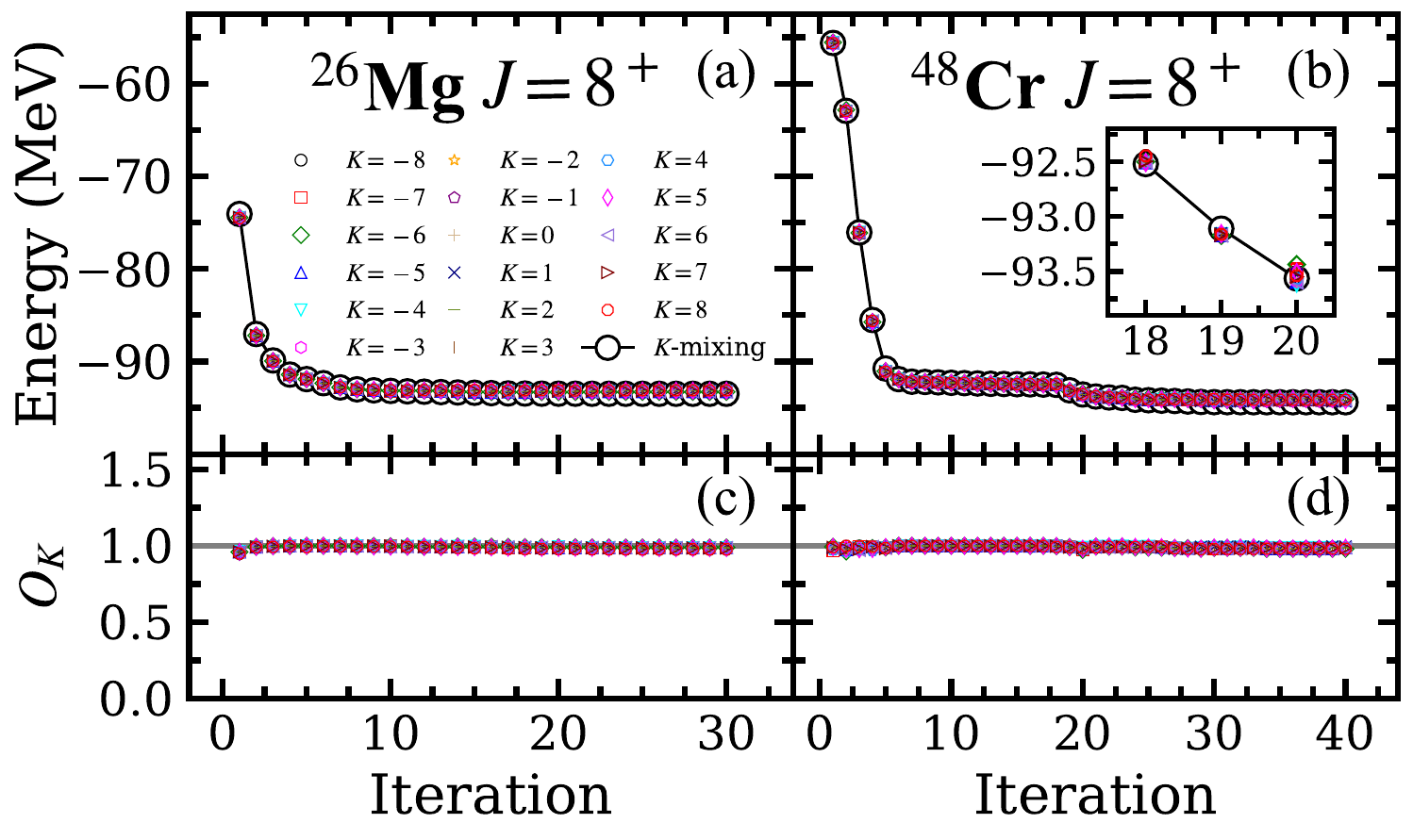}
 \caption{\label{vaprk}(Color online) Energies of the $K$-mixing $|\Psi_{J\pi M1}\rangle$ wave function (black circles) and the corresponding equivalent $P^{J\pi}_{MK}|\Phi'\rangle$ projected states (different colored symbols) as functions of the VAP iteration for the $J^\pi=8^+$ states in (a) $^{26}$Mg and (b) $^{48}$Cr, respectively. The corresponding $O_K$ overlaps are shown in (c) $^{26}$Mg and (d) $^{48}$Cr, respectively. }
\end{figure}

As a final example, we try to identify the validity of Eq. (\ref{pjk}) by considering $|\Phi\rangle$ as a fully symmetry-unrestricted SD.  One can start from a randomly chosen $|\Psi_{J\pi M1}\rangle$ with $|\Phi\rangle$ fully symmetry-unrestricted. Then this $|\Psi_{J\pi M1}\rangle$ can be varied in our VAP calculation, so that the corresponding energy reaches a minimum. Along the path of the VAP iteration, one can harvest a series of different $|\Psi_{J\pi M1}\rangle$ wave functions with the final one fully optimized. Each of them can be put into  Eq. (\ref{rk}) and the $R_K$ can be minimized by varying $|\Phi'\rangle$.  Here, we perform the VAP calculations for $^{26}$Mg in the $sd$ model space and $^{48}$Cr in the $fp$ model space, where the parity projection is unnecessary. The adopted shell model interactions are USDB \cite{usdb} and GXPF1A \cite{gxpf1a}, respectively.  The spins of the calculated wave functions for both nuclei are $J=8$. The calculated results are shown in Fig. \ref{vaprk}. For the $|\Psi_{J\pi M1}\rangle$ at each VAP iteration, whose energy is shown in Fig. \ref{vaprk} (a) or (b), there are $2J+1=17$ $R_K$ quantities with $K=-J, -J+1, \cdots, J$, which can be minimized independently. Then the corresponding 17 maximized $O_K$ quantities for each $|\Psi_{J\pi M1}\rangle$ can be obtained and are shown in Fig. \ref{vaprk} (c) or (d). It is very striking that almost all calculated $O_K$ values are very close to 1. This means at each VAP iteration, the calculated $|\Psi_{J\pi M1}\rangle$ wave function can be equivalently replaced by a single projected state $\hat{P}^{J\pi}_{MK}|\Phi'\rangle$ with $K$ randomly taken. The energies of $\hat{P}^{J\pi}_{MK}|\Phi'\rangle$ are also calculated and are compared with that of the $|\Psi_{J\pi M1}\rangle$ in Fig. \ref{vaprk} (a) and (b). It is seen that the energies of the $\hat{P}^{J\pi}_{MK}|\Phi'\rangle$ projected states perfectly coincide with those of the corresponding $|\Psi_{J\pi M1}\rangle$ wave functions, all the energy differences are within 400 keV. 

%我们之所以可以通过变分操作找到满足16式的SD，其思想与变分方程和薛定谔方程等价相似。即使初始采用一个投影态，通过变分操作，体系会自动的找到能量最小的点。
%One can understand the phenomenon described above from the perspective of nuclear physics. At a given spin $J$, with a $k$-mixing projected wave function, one can obtain the corresponding energy. On the other hand, one can also assume the nucleus is rotating In this process,  the nucleus can always find a minimum with energy very similar to the ones with $k$-mixing projected wave function. 

In summary, we have studied the so called $K$-mixing in the nuclear wave functions which are defined in the laboratory frame of reference. For a given spin and parity  $J^\pi$, one can generate $2J+1$ projected states from a reference state. The nuclear wave function is usually expressed in terms of all these $2J+1$ projected states with different $K$ numbers.  Such form of nuclear wave function have been adopted by almost all nowadays nuclear structure models involving the angular momentum projection. In this work, we have analytically transformed such $K$-mixing wave function into a projected state generated from another complicated reference state $|\phi\rangle$. This means the $K$-mixing in any angular momentum projected wave function can be safely removed. More interestingly, our calculation results indicate that as long as variational calculations are performed, such complicated $|\phi\rangle$ can always be replaced by a simple SD $|\Phi'\rangle$.
%such complicated $|\phi\rangle$ can always be replaced by a simple SD $|\Phi'\rangle$ in all the present calculations. 
This makes the angular momentum projected nuclear wave function to be in a more compact form.

It is well known that the $K$-mixing also exists in the cranked shell model (CSM) \cite{Zeng90}. Such $K$-mixing is caused by the Coriolis force, which is a natural feature of the CSM. However, the Coriolis force only exists in the rotating body-fixed frame of reference. Such inertial force should not exist in the static laboratory frame of reference in which the present wave functions are constructed. Obviously, the presently discussed $K$-mixing is simply originated from the nonzero matrix elements of the Hamiltonian and/or nonzero overlaps between different projected states rather than from the Coriolis force. Thus, conceptually, the $K$-mixing in the angular momentum projected wave function is essentially different from that in the CSM. In this sense, the removal of $K$-mixing of the angular momentum projected wave function does not mean the $K$-mixing in the CSM can be removed.
 
\begin{acknowledgments}
One of the authors (Xiao Lu) extends her gratitude to Shan-Gui Zhou for the helpful discussions. This work is supported by the National Natural Science Foundation of China under Grant Nos. $12347139$ and $11975314$, by the Key Laboratory of Nuclear Data foundation (JCKY2022201C158) and by the Continuous-Support Basic Scientific Research Project.
\end{acknowledgments}


\bibliography{apssamp.bib}
\end{document}